\documentclass[apj]{emulateapj}
\usepackage{apjfonts}
\usepackage{hhline}
\usepackage{amsmath}
\usepackage{graphicx}
\usepackage{natbib}
\usepackage{epstopdf} 
\usepackage{subfigure}
\usepackage{color}
\usepackage{comment}
\usepackage{longtable}

 \font\sevenrm= cmr7 scaled 1000

\newcommand{\ergs}{erg~s$^{\rm -1}$}
\newcommand{\msun}{$M_{\odot}$}       
\newcommand{\OIII}{[O~{\sevenrm III}]}
\newcommand{\OI}{[O~{\sevenrm I}]}

\newcommand{\SII}{[S~{\sevenrm II}]}
\newcommand{\NII}{[N~{\sevenrm II}]}
\newcommand{\FeVII}{[Fe~{\sevenrm VII}]}
\newcommand{\Hb}{H$\beta$}
\newcommand{\Ha}{H$\alpha$}
\newcommand{\kms}{km s$^{-1}$}

\begin{document}
\title{Revisiting the complex kinematics of ionized gas at the central region of NGC 1068: evidence of an additional active galactic nucleus?
}
\author{Jaejin Shin$^{1,2}$}
\author{Jong-Hak Woo$^{2}$\altaffilmark{,4}}
\author{Minjin Kim$^{1}$}
\author{Junfeng Wang$^{3}$}
\affil{
$^1$Department of Astronomy and Atmospheric Sciences, Kyungpook National University, Daegu 41566, Republic of Korea\\
$^2$Astronomy Program, Department of Physics and Astronomy, 
Seoul National University, Seoul, 08826, Republic of Korea\\
$^3$Department of Astronomy, Xiamen University, Xiamen, 361005, China}
\altaffiltext{4}{Author to whom any correspondence should be addressed}

\begin{abstract}
We present a spatially resolved analysis of ionized gas at the nuclear region of the nearby galaxy NGC 1068. 
While NGC 1068 has been known to have gas outflows driven by its active galactic 
nucleus (AGN), more complex kinematical signatures were recently reported, which were inconsistent with a rotation or simple biconical outflows.  
To account for the nature of gas kinematics, we performed a spatially resolved kinematical study, finding a morphologically symmetric pair of approaching and receding gas blobs in the northeast region. 
The midpoint of the two blobs is located at a distance of 180 pc from the nucleus in the projected plane. The ionized gas at the midpoint shows zero velocity and high velocity dispersion, 
which are characteristics of an outflow-launching position, as the two sides of a bicone, i.e., approaching and receding outflows are superposed on the line of sight, leading to no 
velocity shift but high velocity dispersion. We investigate the potential scenario of an additional AGN based on a multiwavelength data set. 
While there are other possibilities, { i.e., X-ray binary or supernova shock}, the results from optical spectropolarimetry analysis are consistent with the presence of an additional AGN, which likely originates from a minor merger. 

\end{abstract}

\keywords{
     galaxies: active ---
     galaxies: individual (NGC 1068) ---  
     ISM: jets and outflows ---
     techniques: imaging spectroscopy}

\section{INTRODUCTION} \label{section:intro}
NGC 1068 at a distance of 14.4 Mpc \citep{Bland-Hawthorn1997}
is a prototype Seyfert 2 galaxy with a well-studied active galactic nucleus (AGN). The first detection of the 
hidden broad-line region (BLR) was reported based on spectropolarimetric observations \citep[e.g.,][]{Antonucci1985}. 
Since then, it has been widely accepted that the dichotomy of Seyfert 1 and Seyfert 2 galaxies is due to an 
orientation effect caused by a dusty molecular torus surrounding the BLR.
The presence of molecular tori has been observationally confirmed by high spatial 
resolution observations \citep[e.g.,][]{Garcia-Burillo2016,Imanishi2016,Imanishi2018,Combes2019}.
The mass of the central black hole was reliably measured as M$_{\rm BH}$ =  0.8--1.7$\times 10^{7} \rm \ M_{\odot}$
based on spatially resolved maser kinematics \citep{Greenhill1996,Hure2002,Lodato2003,Impellizzeri2019},
while the bolometric luminosity was determined as L$_{\rm Bol}$ =  0.4--4.7$\times 10^{45}\ \rm erg \rm s^{-1}$
in various ways using multiwavelength data \citep[][and references therein]{Gravity2020}. 
The corresponding Eddington ratio ranges from 0.19 to 4.7, indicating a high accretion rate.

One intriguing feature of NGC 1068 is that ionized gas kinematics shows complexity 
in the central region (i.e., < 500 pc) as reported by \cite{Walker1968} 
and subsequent studies. Using various spectroscopic data, including HST observations, detailed analysis has been 
conducted to constrain the origin of the gas kinematics. Biconical gas outflows have mainly been interpreted 
as being driven by the central AGN and manifest as an approaching gas blob in the northeast (NE) region and a receding gas 
blob in the southwest (SW) region from the nucleus
\citep{Cecil1990,Arribas1996,Axon1998,Crenshaw2000a,Cecil2002,Das2006,Gerssen2006}. 

In contrast, there were reports of an additional receding (redshifted) gas blob, which was detected at 
2.5--4.5\arcsec\ (i.e., 180--320 pc) NE of the nucleus \citep[e.g.,][]{Axon1998, Cecil2002}. Because this gas blob 
disagrees with the trend of the well-known biconical outflows (i.e., approaching in NE), more complex 
mechanisms are likely to be responsible for the change in the velocity sign in the NE region.  
While the receding gas blob in the NE region was claimed to be the result of 1) the lateral expansion of radio 
jets \citep{Axon1998,Cecil2002}, or 2) escaped radiation through a patchy dust torus or scattered radiation 
\citep{Das2006}, the origin of the complex gas kinematics is yet to be clearly understood.

To constrain the origin of the complex gas kinematics at the nuclear region, we performed 
spatially resolved spectroscopic analysis by utilizing the VLT/MUSE data. We describe the data and 
analysis in Section 2. The main results and discussion are followed in Section 3 and 4. Conclusions are presented 
in Section 5. We adopt a cosmology with
$H_{\rm 0}=  70$ km  s$^{-1}$ Mpc$^{-1}$, $\Omega_{\Lambda}= 0.7$ and $\Omega_{\rm m}= 0.3$.  \\

\section{Data and analysis}\label{section:Sample}
\subsection{Data}\label{section:Sample}

 NGC 1068 was observed with the VLT/MUSE as a part of the MAGNUM survey (094.B-0321(A) PI: A. Marconi),
 which covered a 1\arcmin $\times$1\arcmin\ (i.e., 4.3 kpc$\times$4.3 kpc) field of view (FOV).
\cite{Mingozzi2019} analyzed the MUSE data to investigate the properties of outflowing gas 
(e.g., density and ionization parameter), while the kinematics of ionized gas, particularly at the central 1 kpc scale, 
was not presented in their work. In this paper, we focus on the central region and present the detailed gas kinematics. 
The observation consists of 
12 exposures, the exposure time of which was 500 s for four exposures (observed on 2014 Oct 6) and 100 s 
for eight exposures (observed on 2014 Dec 1). For this work, we only utilized the eight 100 s exposures since 
\OIII\ and \Ha\ are saturated at a few central pixels in the 500 s exposures. The data were retrieved from 
the ESO archive and reduced using the standard reduction pipeline of the VLT instruments, 
ESOREFLEX \citep{Freudling2013}. The seeing of the individual exposures ranged 
from 0.81 to 0.98\arcsec, corresponding to $\sim$58--71 pc at the distance of NGC 1068.

\begin{figure*}{}
\centering
\includegraphics[width =  1\textwidth]{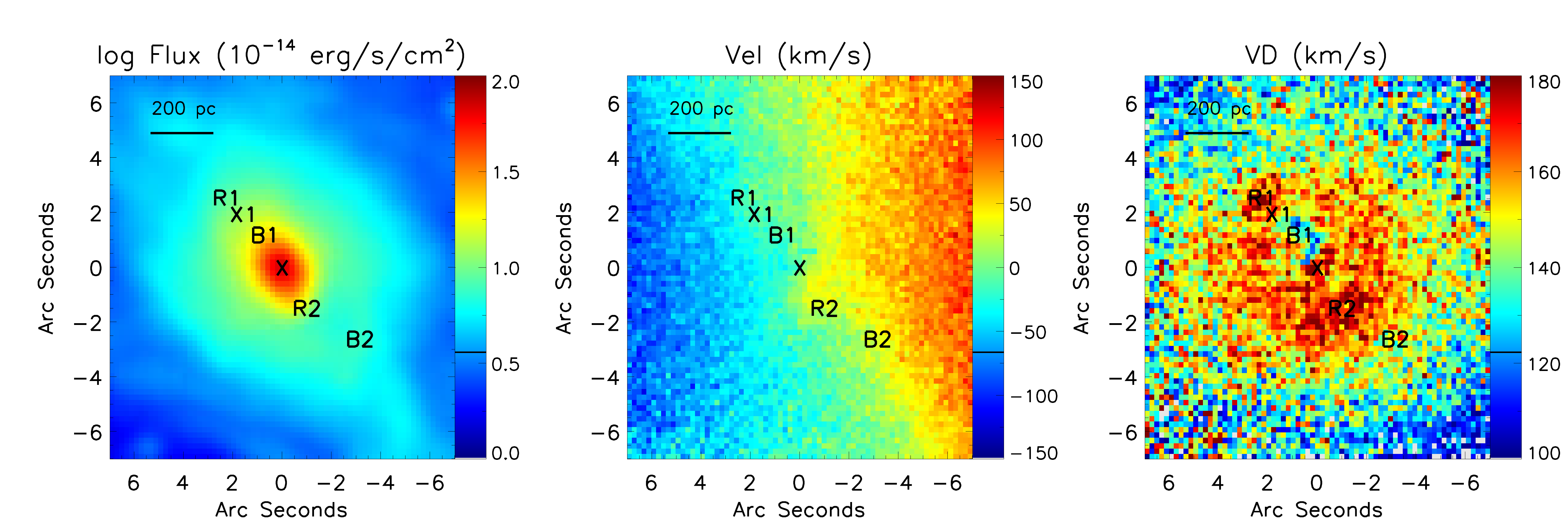}
\includegraphics[width =  1\textwidth]{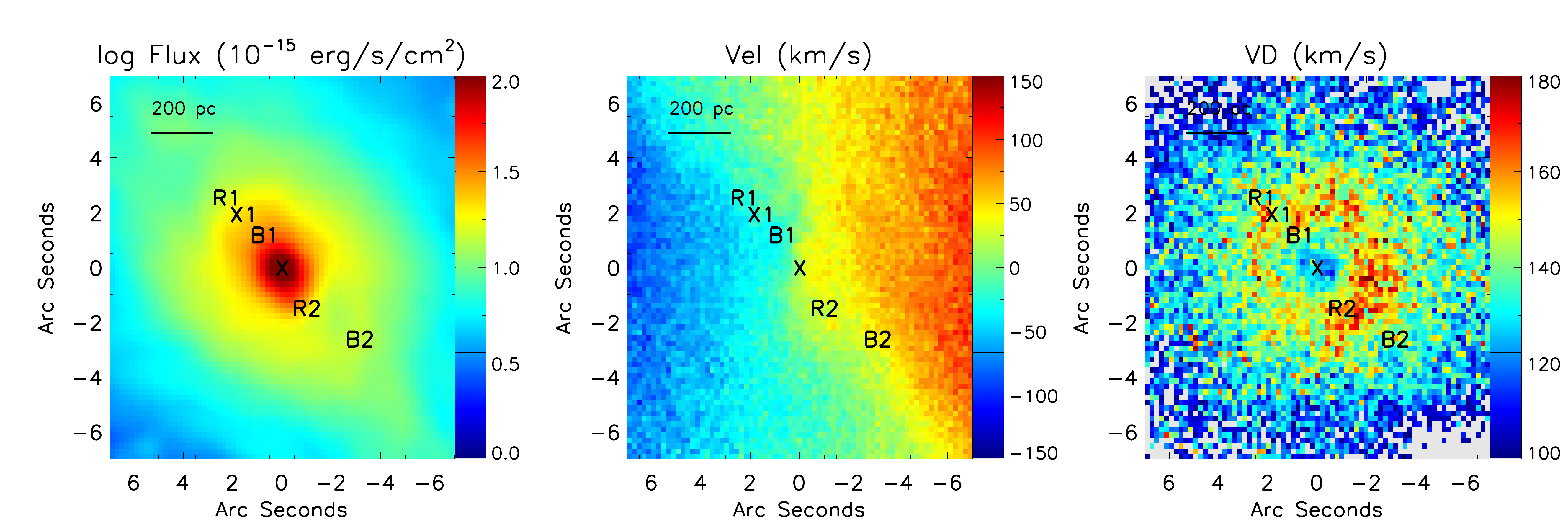}
\caption{
Flux, velocity, and velocity dispersion maps of the stellar component analyzed in two wavelength ranges, 
1) 4800-6800\AA\ (upper) and 2) 8400-8800\AA\ (lower).
The center of each map is defined with the the highest flux point (X) in the stellar continuum flux map.
Other marks (e.g., R1) will be described in Section 3.2.
\\
\label{fig:allspec1}}
\end{figure*} 

\begin{figure*}{}
\centering
\includegraphics[width =  1\textwidth]{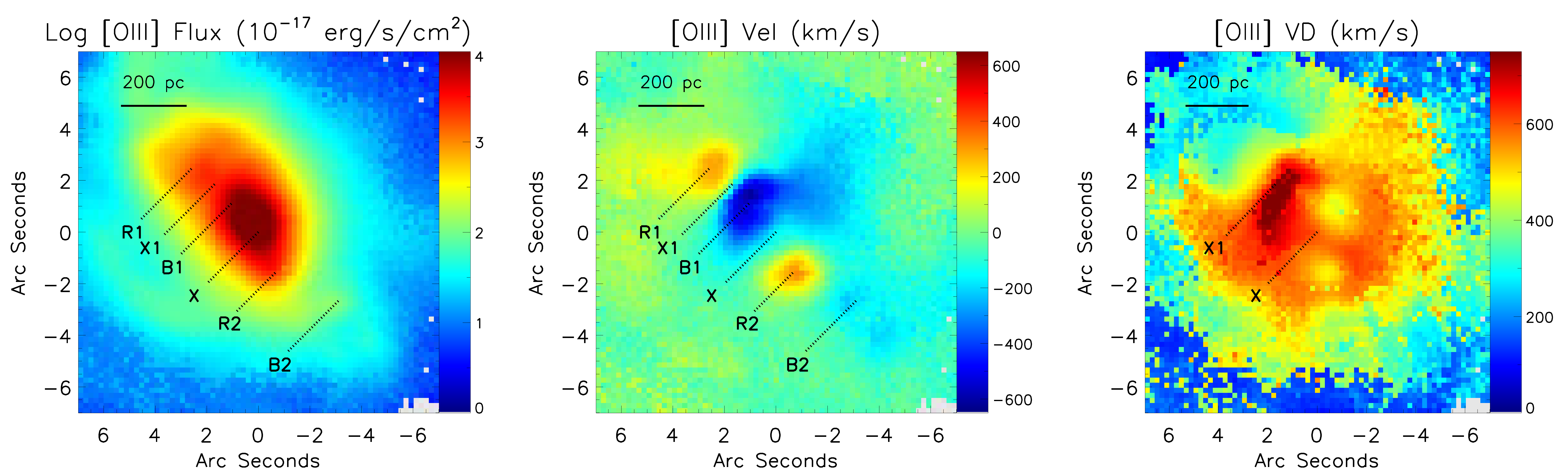}
\includegraphics[width =  1\textwidth]{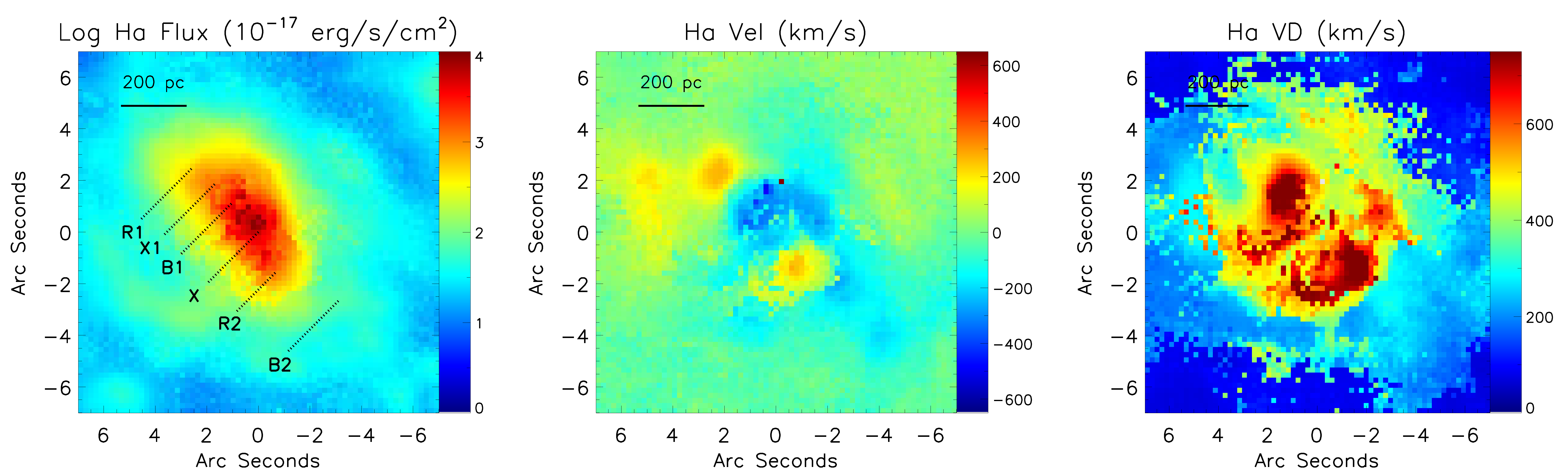}
\includegraphics[width =  1\textwidth]{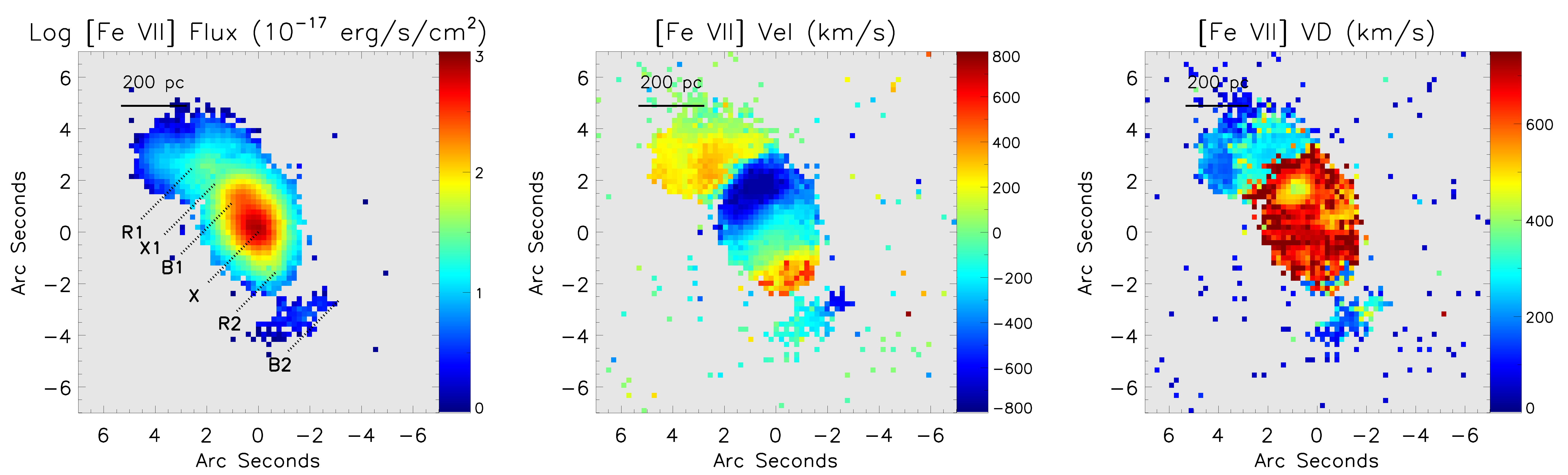}
\caption{
Flux, velocity, and velocity dispersion maps of \OIII\ (top), \Ha\ (middle), and \FeVII\ (bottom). Four high-velocity gas blobs (i.e., R1, R2, B1, and B2)
and a region with zero velocity and high velocity dispersion (X1) are marked based on the velocity and velocity dispersion maps of \OIII. 
The nucleus of NGC 1068 is indicated as X. The center of each map is defined with the the highest flux point in the stellar continuum flux map as same as Figure~1.
\\
\label{fig:allspec1}}
\end{figure*} 

\subsection{Analysis}\label{section:Sample}
We performed a spectral fitting analysis by focusing on the central 
14\arcsec$\times$14\arcsec\ (or 1 kpc$\times$1 kpc) region.
While the MUSE data covered several kpc scales as described in Section 2.1, the complex gas kinematics was reported within the central 
$\sim$ 5\arcsec\ (or $\sim$350 pc) region, where ionized gas showed high velocity as well as high velocity dispersion 
\citep[i.e., $V_{\rm \OIII}$ > 1000 \kms\ and $\sigma_{\rm \OIII}$ >1000 \kms, e.g.,][]{Cecil2002,Das2006}. 
Outside the central $\sim$ 5\arcsec\ region, gas velocity and velocity dispersion are typically less than 100 \kms, 
which are comparable to those of the stellar component of NGC 1068 \citep[e.g.,][]{Emsellem2006}. 
We checked the whole FOV of the MUSE data (1\arcmin$\times$1\arcmin) and 
confirmed that gas kinematics in large scales generally follows the rotation of the galaxy 
without showing a strong outflow signature (i.e., high velocity). 
We analyzed optical emission lines, i.e., 
\OIII\ $\lambda$5007, \Ha, and \FeVII\ $\lambda$6087 to trace gas kinematics
and stellar absorption lines to investigate stellar kinematics, as extensively performed in our previous works \citep[e.g.,][]{Bae2014,Karouzos2016a, Karouzos2016b, Woo2016,Bae2017,Woo2017,
Kang2018, Luo2019,Shin2019b}.

First, we fitted and subtracted the stellar continuum on a spectral range of 4800-6800\AA, using the pPXF code \citep{Cappellari2017} 
with 47 ages (from 60 Myr to 12.6 Gyr) and solar metallicity of the E-MILES template \citep{Vazdekis2016}. 
During the fitting, we masked every visible emission line as well as the Na I $\lambda$$\lambda$5890,5896 (Na D) doublet absorption line, which traces neutral 
gas outflows \citep[e.g.,][]{Bae2018} to avoid any possible contamination in the stellar continuum fitting. 
Then, 
we fitted emission lines (i.e., \OIII\ \& \Ha), with the MPFIT package \citep{Markwardt2009}. 
To reproduce the observed emission lines, we adopted multiple Gaussians (up to three), whose amplitude-to-noise ratios 
are all larger than 3. The number of Gaussians was determined with the reduced chi-square of the fitting results. In the fitting, 
we tied the velocity and velocity dispersion for the (1) \OIII\ doublet and (2) \Ha\ and \NII\ doublet, respectively.  
We also fixed flux ratios of \NII\ $\lambda \lambda$6548,6584 and \OIII\ $\lambda \lambda$4959,5007 doublets as 3 and 2.98.
Note that we used a consistent method as adopted by \cite{Shin2019b}, who analyzed the MUSE data of NGC 5728, 
except for the number of Gaussians (up to two in \citealt{Shin2019b}).

From the best-fit model, we measured the flux, velocity, and velocity dispersion of each line from its whole line profile
(i.e., flux-weighted line properties). Different from this work \citep[and also][]{Emsellem2006}, each Gaussian 
component was investigated to understand individual gas components in several previous works of 
NGC 1068 \citep{Cecil2002,Das2006,Gerssen2006}. They divided the multiple Gaussian components 
based on line flux \citep[i.e., highest to lowest;][]{Das2006} or velocity \citep[redshifted or blueshifted;][]{Cecil2002}. 
We tried to decompose the individual Gaussian components based on the same criteria (i.e., flux and velocity), while 
the decomposed components in 2D space show complex distributions, which is very difficult to interpret. 
Therefore, we measured the flux-weighted line properties to investigate the overall (or dominant) gas kinematics for the first step. 

Second, we fitted the stellar continuum in the wavelength of 8400-8800\AA\ with the same method as used for 4800-6800\AA.
Seyfert 2 galaxies generally show strong stellar absorption lines (i.e., Mgb, Fe5270, and Fe5335) in the optical wavelength
\citep[e.g., NGC 5728,][]{Shin2019b}. This is also true for NGC 1068 except 
for its central $\sim$2\arcsec\ $\times$2\arcsec\ region, where a strong nonstellar continuum is present, leading to unreliable 
fitting results of the stellar component \citep[see Figure~6 and 7 of][]{Gerssen2006}. The MUSE data also showed a
strong nonstellar continuum in the same central region. Therefore, we used another wavelength window, 8400-8800\AA, 
where the CaII triplet is clearly detected representing stellar kinematics.

Even in this wavelength range, we also detected a nonstellar continuum and the features of emission lines, which are
associated with e.g., [Cl II] $\lambda$8579, Pa14 $\lambda$8598, [Fe II] $\lambda$8617, and [N I] $\lambda$8629. To avoid any contamination, 
we masked the wavelength window of 8555-8650\AA\ during the fitting and measured the stellar velocity and velocity dispersion.  
Overall, the stellar model fitting results are better in the CaII triplet region than in the Mgb-Fe region, indicating that 
the measured stellar kinematics from 8400-8800\AA\ could be more reliable than that from 4800-6800\AA.\\

\section{Result} \label{section:result}

\subsection{Stellar component}\label{astrometry correction}
In Figure~1, we present the maps of the flux, velocity, and velocity dispersion of the stellar component analyzed in the 
two wavelength ranges, (1) 4800-6800\AA, and (2) 8400-8800\AA.  
Throughout the paper, we consider the highest flux point in the stellar continuum flux map as the center (0, 0) of NGC 1068.
Without any astrometry correction, the center is offset by only $\sim$0.1\arcsec\ (i.e., the half of the pixel size of the MUSE data) 
from the position of the peak of X-ray emission at the nucleus of NGC 1068 (RA=  2:42.40.71, Dec=  --00:00:47.7, \citealt{Young2001}).
We find that both of them commonly show a radially decreasing flux and a rotation pattern. We also find that the mean 
difference between the two velocity maps is 12.5$\pm$20 \kms. This means that both velocity maps can be 
used as a reference for the systemic velocity. 

In the case of the velocity dispersion maps, however, there is a significant discrepancy (the right panels of Figure~1). 
Using the 4800-6800\AA\ range, we find high velocity dispersion along the NE-SW direction. In contrast, when the 8400-8800\AA\ 
region was used, we detect a trend of $\sigma$ drop at the central circular region within a radius of $\sim1\arcsec$. 
The central $\sigma$ drop was detected in a number of galaxies including NGC 1068 \citep{Ensellem2004,Emsellem2006,
Falcon-Barroso2006,Gerssen2006,Comeron2008,Kang2013} and its origin has been discussed as recent star formation at 
the nuclear region \cite[e.g.,][]{Ensellem2001,Comeron2008,SB2012}.
We note that the reason for the high velocity dispersion along the NE--SW in the former one (the top-right panel of Figure~1) would 
be mainly due to the unreliable fitting result of the stellar continuum at the central region as described in Section 2.
As also presented in \cite{Gerssen2006}, the stellar continuum fitting is unreliable at the region with high velocity dispersion.\\

\begin{figure*}{}
\centering
\includegraphics[width =  0.43\textwidth]{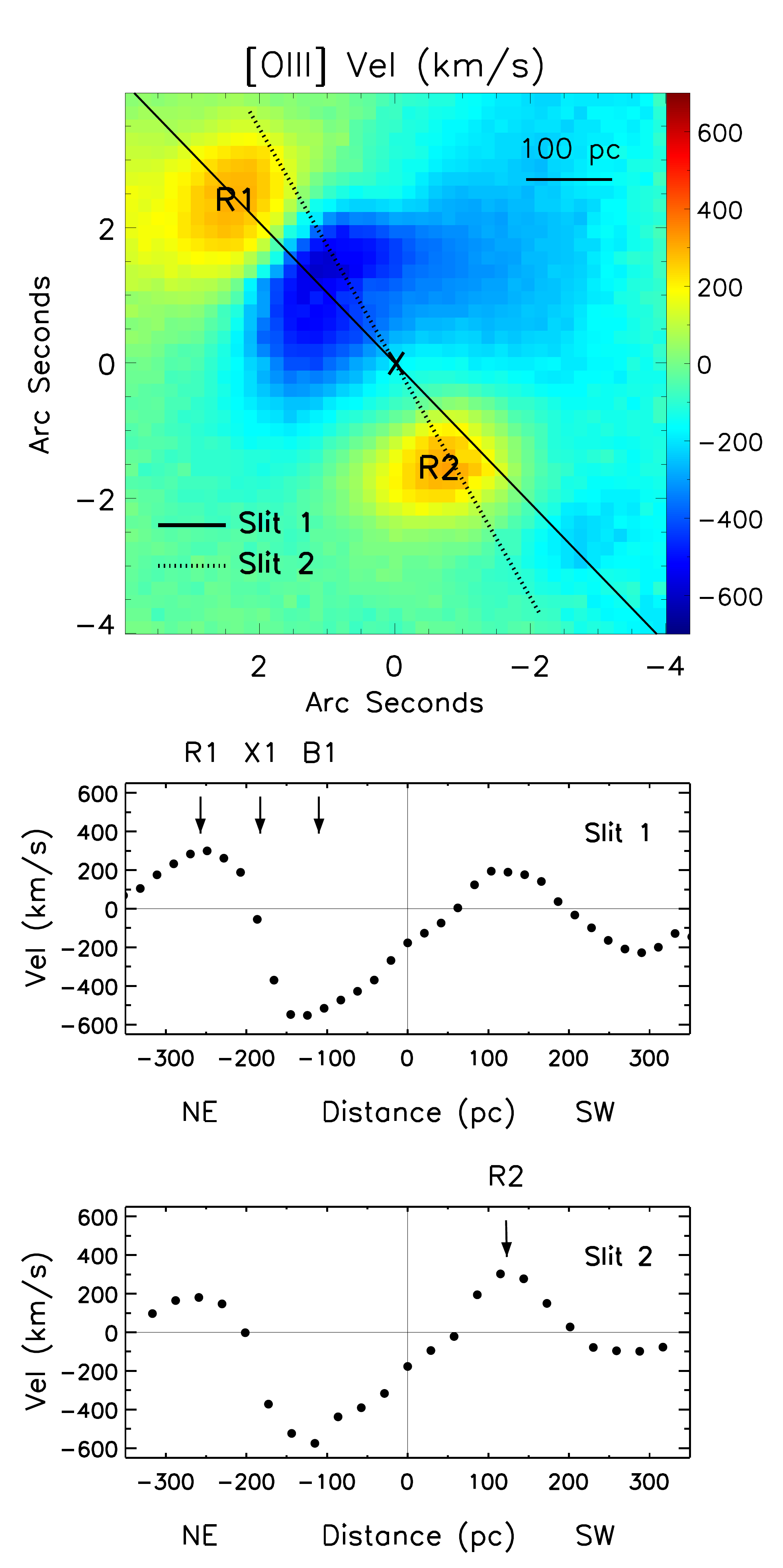}
\includegraphics[width =  0.43\textwidth]{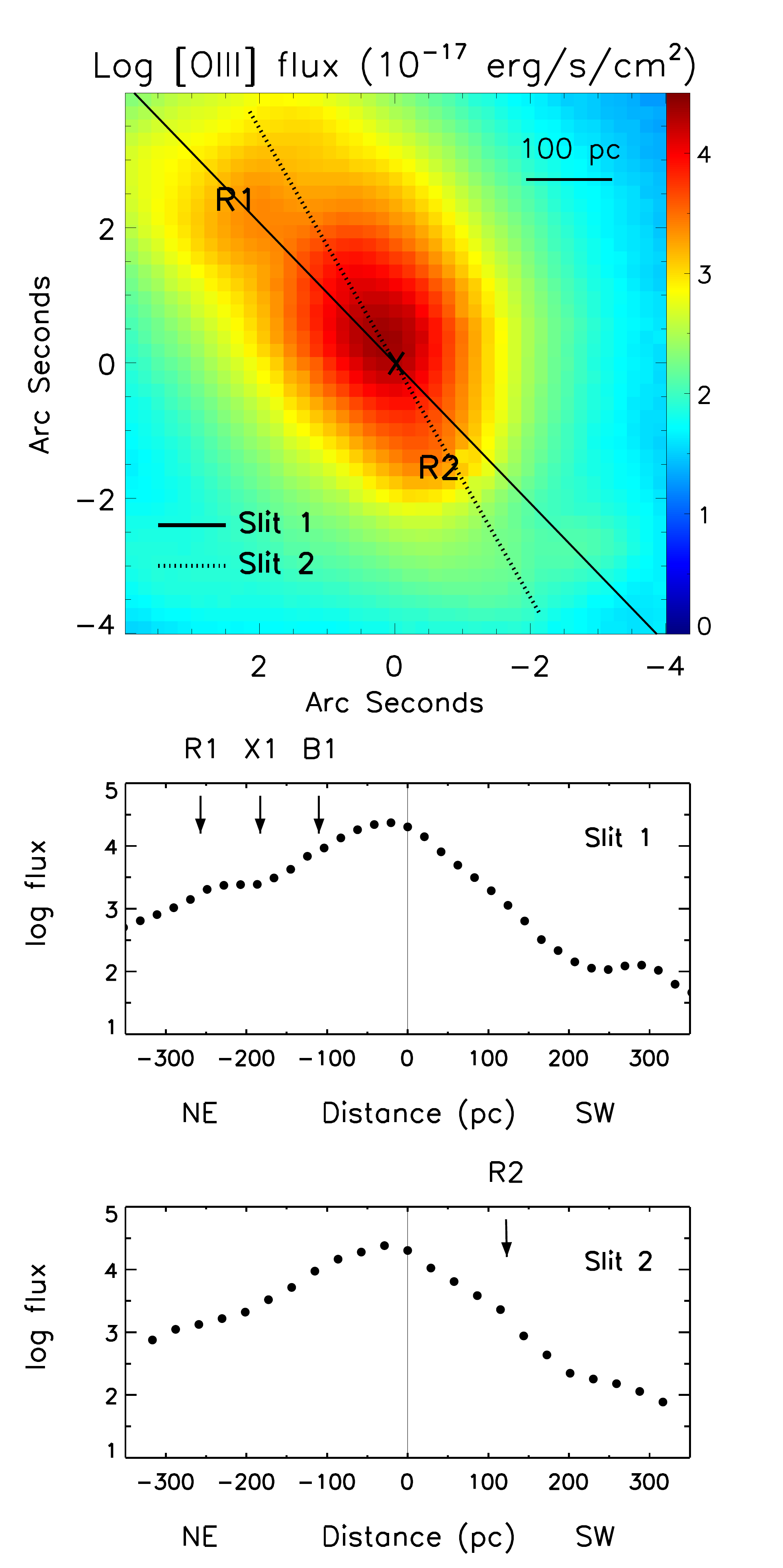}
\caption{
Top: velocity and flux map of \OIII\ with two pseudo-slits, which (1) through the nucleus of NGC 1068 and R1 (slit 1, solid line, PA= 44$\arcdeg$) 
and (2) along the orientation of the radio jet (slit 2, dotted line, PA= 30$\arcdeg$, \citealt{Wilson1983}). 
One-dimensional velocity { and flux} distributions along slit 1 (middle) and 2 (bottom) are shown with the positions of high-velocity gas blobs (i.e., R1).
Note that the center of R2 is not exactly on slit 2, while we simply mark R2 at the position of the foot of perpendicular 
from the center of R2 to the slit 2.
\\
\label{fig:allspec1}}
\end{figure*}

\subsection{Ionized gas}\label{astrometry correction}
We present the flux, velocity, and velocity dispersion maps of the two major optical emission lines, \OIII\ and \Ha, as well as a coronal emission line, \FeVII\ in Figure~2.
  \OIII\ and \Ha\ represent the ionized gas kinematics in the NLR, while \FeVII\ with a
 high ionization potential (99.1 ev) traces the coronal line region \citep[CLR; e.g.,][]{Rodriguez-Ardila2006,Rodriguez-Ardila2020}.
In the velocity maps, we use relative velocity (i.e., $V_{\rm gas}$--$V_{\rm \ast}$) at each spaxel in order to show 
the relative motion of gas with respect to stellar component, for which velocity is measured from stellar absorption line in the 8400-8800\AA\ range (bottom-middle panel of Figure~1). Because stellar velocity is much smaller than gas velocity,
we find no significant change in the velocity map with/without subtracting stellar velocity. 

First, the flux maps show an elongated distribution along the NE-to-SW direction,
which represents an ionizing cone as discussed in the previous studies \citep[e.g.,][]{Capetti1997}. 

Second, the velocity maps present a complex structure, with a wide range of line-of-sight velocity from --700 \kms\ to +400 \kms). The much higher gas velocity than that stellar velocity indicates 
that gas kinematics is mainly governed by non gravitational effect, i.e., AGN-driven outflows, 
instead of the gravitational potential of the host galaxy. All velocity maps 
show a common kinematic structure, by presenting four high-velocity gas blobs (i.e., $\vert V\vert$ > 200 \kms): one pair of blueshifted (B1) and redshifted (R1) blobs in the NE direction, and another pair of blueshifted (B2) and redshifted blobs (R2) 
in the SW direction as marked in the \OIII\ velocity map. Interestingly, R1 and B1 show morphologically 
similar tail-like structures, which are elongated roughly in the horizontal direction, implying that R1 and B1 are physically 
connected as a pair.  
The four gas blobs are also detected in the \FeVII\ velocity map,
suggesting that the high-velocity gas blobs are 
driven by AGN rather than star formation because \FeVII\  has a high ionization potential, i.e., 99.1 eV. 
The positions of these gas blobs are slightly different depending on the tracer. For example, there is a $\sim$ 0.2\arcsec\ offset of the center of R1 between the \OIII\ and \Ha\ velocity maps. The spatial offset is presumably due to the difference in ionizing sources.
While \Ha\ (and also \Hb) is ionized by AGN as well as star formation, \OIII\ and \FeVII\ are mainly ionized by AGN. 
In addition, the possible uncertainty of the \FeVII\ velocity, which can be much larger than that of \Ha\ and \OIII\ 
because of the lower flux of \FeVII\ (see Figure~2), can be partly responsible for the offset. 
Because the exact position of these gas blobs is not the main interest for our analysis, we determined the location of gas blobs using the \OIII\ velocity map. 

The high-velocity gas blobs are spatially resolved in the HST narrowband image \citep{Capetti1997}. 
For example, the gas blob B1 is resolved into a couple of smaller blobs. 
Without velocity information, however, it is difficult to constrain how the substructure of B1 detected in the narrowband image is related to the outflows.
 Previous studies of gas kinematics based on the HST/STIS data showed consistent features compared to those presented in this work \citep{Cecil2002,Das2006}. For example, a cloud 
located at the center of B1 also showed negative velocities, $\sim$600 \kms\ (see Figure~3 of \citealt{Cecil2002}).

The velocity structure presented in this work is very different compared to the biconical outflows in Seyfert galaxies, which typically show a pair of blueshifted and redshifted gas blobs (e.g., NGC 5728; \citealt{Durre2019}; \citealt{Shin2019b}). 
While the previous studies treated B1 and R2 as approaching and receding gas outflows, respectively, driven by the AGN at the nucleus of NGC 1068 \citep[e.g.,][]{Das2006}, there are apparently additional gas blobs (R1 and B2), whose gas kinematics is 
inconsistent with a single biconical outflows. 

To better understand the intriguing kinematical structure, we present one-dimensional velocity distributions by locating two pseudo-slits  in the \OIII\ velocity map: 1) one through the nucleus of NGC 1068 and R1 (slit 1, solid line, PA= 44$\arcdeg$) and 2) another along the orientation of the radio jet (slit 2, dotted line, PA= 30$\arcdeg$, \citealt{Wilson1983}).
As shown in Figure~3, the position angle of Slit 1 is tilted 
by $\sim$14$\arcdeg$ from the orientation of a radio jet, which has PA=30$\arcdeg$ \citep{Wilson1983}. 
We find a dramatic variation of gas velocity along Slit 1 (i.e, B1 to R1), which 
indicates the change of outflow direction at the radius of $\sim$ --190 pc (and also $\sim$200 pc). 
The change of gas velocity along Slit 2 (i.e., jet direction through R2) is slightly less, but clearly shows
a change in velocity sign. 
Moreover, we detect a significant flux bump around R1 (middle right panel of Figure~3).
While it is difficult to quantify the additional flux, the \OIII\ flux around R1 is somewhat increased ($<$$\sim$0.1-0.2 dex), particularly 
compared to the flux distribution of Slit 2. In contrast, there is no significant flux excess in R2.
Based on these results, we interpret that the kinematical signature of R1 is not related to the radio jet, 
but caused by the flux contribution from another source at X1.

Third, the velocity dispersion maps 
also show a complex kinematical structure.  
Within 5\arcsec\ (i.e., $\sim$ 350 pc) from the nucleus, the velocity dispersion (>300 \kms) is significantly higher 
than stellar velocity dispersion ($\sim$150 \kms), indicating the strong nongravitational kinematics of ionized gas. 
We find that the velocity dispersion is even higher (up to 800 \kms) at the midpoint between B1 and R1 (marked as X1 in Figure~2). 
In contrast, the gas velocity is almost zero at X1. 
Because high velocity dispersion and zero velocity are typically detected at the launching point of gas outflows in nearby AGNs 
(e.g., NGC 5728; \citealt{Durre2019, Shin2019b}), these features suggest that the R1 and B1 gas blobs represent outflows, 
which are launched at the position of X1. \\
 
\begin{figure*}{}
\centering
\includegraphics[width =  0.9\textwidth]{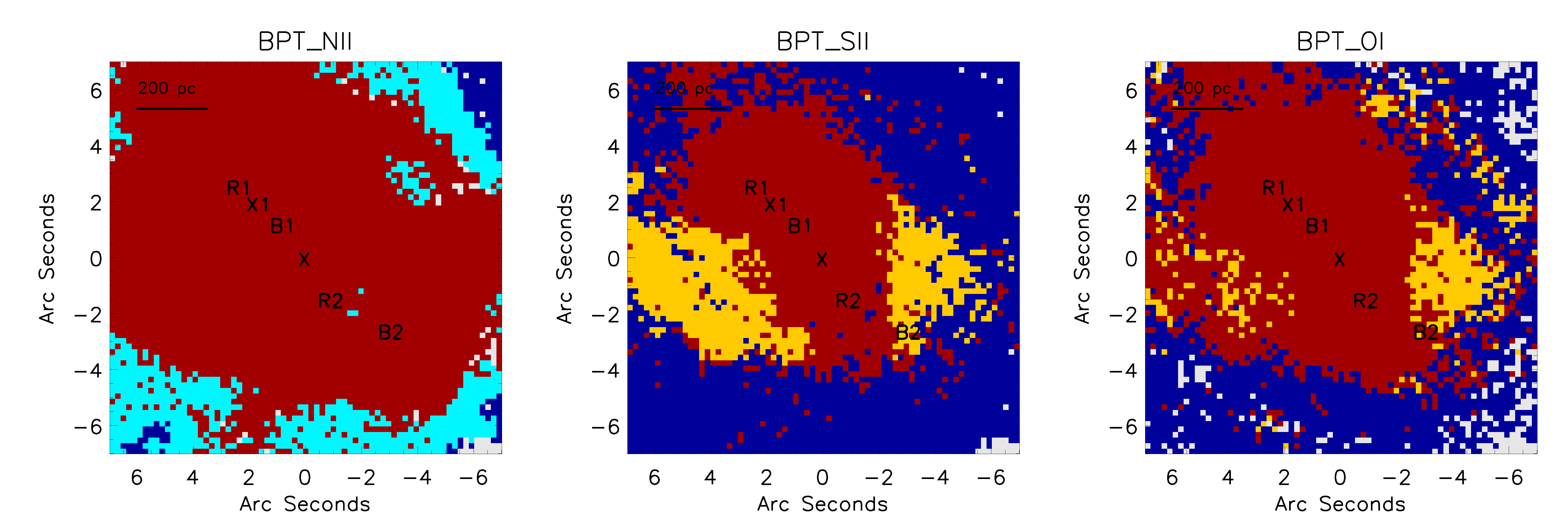}
\caption{
BPT morphology maps with three diagnostics (\NII, \SII, and \OI). Red, yellow, cyan, and blue indicate AGN, 
LINER, composite region, and star-forming region, respectively. Marks are the same as in the top-middle panel of Figure~2.\\
\label{fig:allspec1}}
\end{figure*}

 \subsection{Emission-line flux ratios}\label{astrometry correction}
 In order to understand the ionization mechanism,
we investigate emission-line flux ratios and identify major ionizing source (i.e., AGN, star-forming, 
low-ionization nuclear emission-line region, (LINER), or composite region), using the BPT 
diagrams with three diagnoses \citep[\NII, \SII, and \OI, e.g.,][]{Baldwin1981,Kewley2006}.
We confirm that AGN and LINER are the major ionizing sources within 5\arcsec\ (or $\sim$ 350 pc), while
star formation ionizes gas in the outskirts of the FOV as previously reported \citep[see Figure 4;][]{DAgostino2019,Mingozzi2019}. 
The high-velocity gas blobs are mainly ionized by AGN, suggesting that outflows in these blobs are likely driven by AGN. 

In Figure 5, we present the \OIII/\Hb\ ratio map to trace ionization parameter. As expected from the BPT maps, the ratio is high (>$\sim$5) in the ionized region, where AGN is the main ionizing source. However, the \OIII/\Hb\ ratio ranges from $\sim$ 5 to $\sim$ 10,
and the R1 and B1 blobs show relatively high  \OIII/\Hb\ ratios.
We find two interesting features in the ratio map. First, the highest ratio is detected in the gas blob B1 (i.e., $\sim$1\arcsec\ NE from the position X), instead of the nucleus. By reporting this feature, previous studies discussed that shocks from the interaction between the ratio jet and ISM play as an additional ionizing source in elevating the \OIII/\Hb\ ratio in the NE region 
\citep[e.g.,][]{Kraemer1998,Kraemer2000,Cecil2002}.  
 Second, the region with a high \OIII/\Hb\ ratio (i.e. >$\sim$5) has a biconical shape centered at X1 
(see red dashed lines in Figure~5), suggesting that B1 and R1 are a pair of gas blobs, which are launched at X1 as discussed in the previous section.\\

 \begin{figure}{}
\centering
\includegraphics[width =  0.45\textwidth]{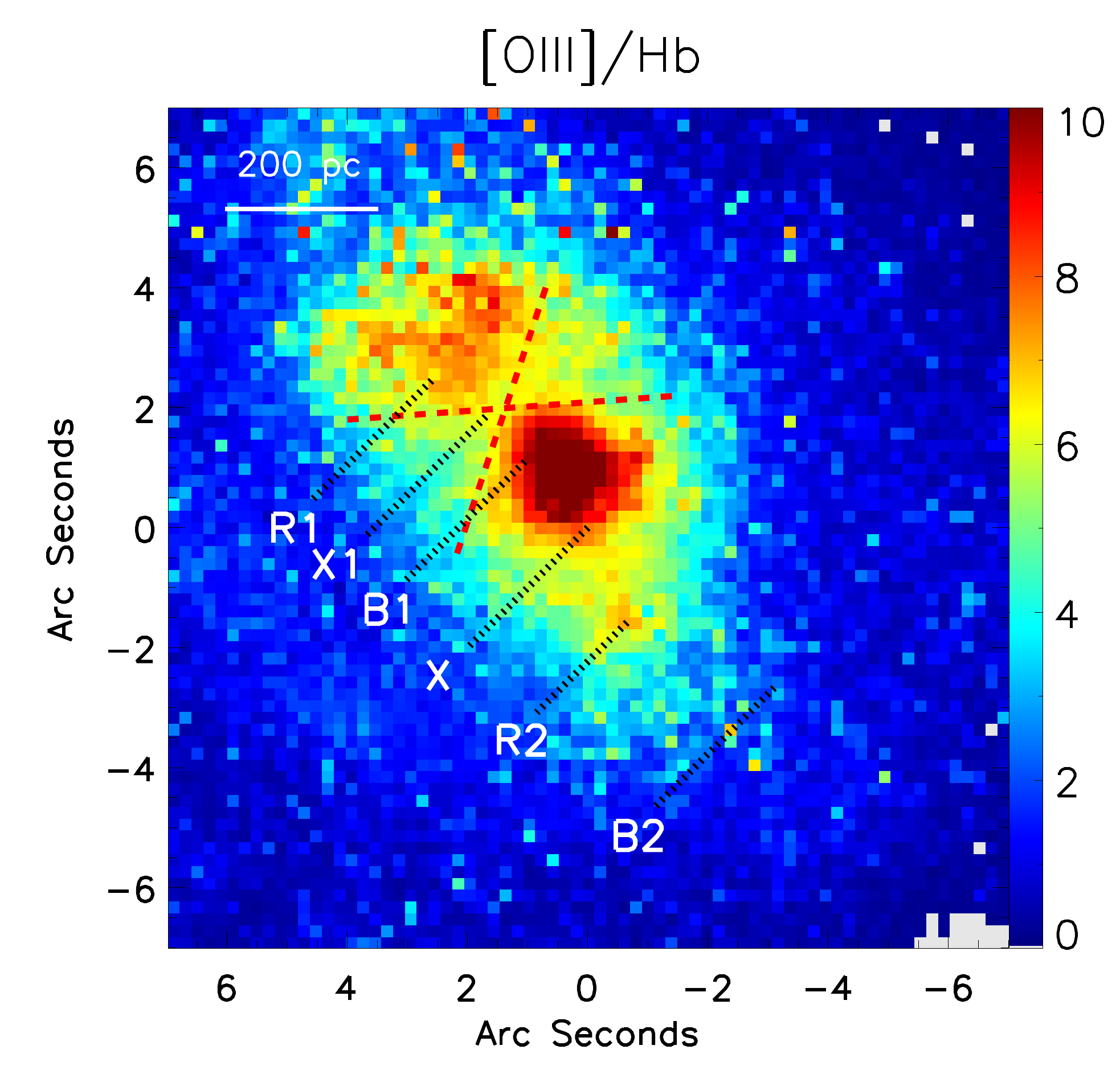}
\caption{
\OIII/\Hb\ ratio map. Red dashed lines represent a bicone shape centered at X1. Marks are the same as in the top-middle panel of Figure~2. \\
\label{fig:allspec1}}
\end{figure}

\section{Discussion}\label{Discussion}

\subsection{The origin of intriguing kinematics}\label{Discussion}

Based on the spatially resolved analysis of the central region of NGC 1068,
we present the complex kinematical signature of ionized gas,
which is inconsistent with either the host galaxy gravitational potential or a simple biconical outflow scenario. 
We find a pair of two distinct gas blobs (i.e., R1 and B1),
which show a kinematically similar morphology in the gas velocity map and share a consistent radial velocity structure with a similar
increasing and decreasing trend. Based on the kinemorphological symmetry, we interpret that the two blobs represent
a pair of receding and approaching gas blobs, which are centered at a launching position (X1) at a projected distance of 180 
pc from the galaxy center.

While the gas blob B1 is interpreted to be the approaching side of outflows, which are launched from the 
nucleus \citep[e.g.,][]{Cecil2002,Das2006}, the gas blob R1 is moving in the opposite
direction compared to B1. To explain the origin of R1, two scenarios have been suggested: 
(1) the lateral expansion of radio jets \citep[e.g.,][]{Axon1998,Cecil2002} and 
(2) escaped radiation through a patch torus or scattered radiation \citep{Das2006}. These scenarios 
may explain the opposite orientation of R1 compared to B1 (see, e.g., Figure~6 of \citealt{Cecil2002}).

However, neither scenario is acceptable because such a change of velocity sign in outflows has not been 
reported in any other AGNs observed with gas outflows and radio jets.
For example, in the cases of well-studied NGC 4151 and NGC 5728, the velocity structure of gas outflows is 
consistent with biconical outflows, manifesting as approaching gas in one side and receding gas in the other side, 
without changing the velocity sign in one direction \citep[e.g.,][]{Das2005, Durre2019, Shin2019b}.

Moreover, we find that the orientation from the nucleus (X) to the position X1 (i.e., PA= 44$\arcdeg$) is tilted by $\sim$14$\arcdeg$
from the PA of the radio jet, which has PA= 30$\arcdeg$. If the lateral expansion of radio jets and/or escaped radiation is the 
origin of R1, it is natural to expect a similar gas blob, which is laterally expanded from the jet and located in the opposite direction 
(i.e., PA=  16$\arcdeg$; see the slit position in Figure 3). In other words, a symmetric gas blob would be located on the other side of the jet. 
{ In contrast, we find no structure in that position (at $\Delta$RA=  $\sim$0\arcsec, $\Delta$Dec= $\sim$3\arcsec) while it is possible that the lack of ionized gas in that location may explain why the lateral expansion or escaped radiation is morphologically asymmetric (top-right panel of Figure~3).} In addition, it was pointed out that the radio jets are not energetic enough to drive strong gas outflows \citep{Das2006}.  
Therefore, the two aforementioned scenarios cannot fully explain the observed flux and velocity structures at around R1, 
implying that the AGN at the nucleus of NGC 1068 may not be responsible for the kinematics of the gas blob R1.

We propose a new scenario where there is an additional (second) mass-accreting black hole at around the position X1, 
and the outflows driven by the 2nd AGN at X1 are manifested by a pair of receding (R1) and approaching (B1) gas blobs. 
We present several supporting evidence. First, as presented in Section 3.2, ionized gas at X1 shows zero velocity 
and high velocity dispersion, which are typical observational characteristics of an outflow-launching point in nearby Seyfert galaxies 
\citep[e.g., NGC 5728;][]{Durre2019,Shin2019b}. This trend is due to the fact that the two sides (i.e., receding and approaching) 
of the gas outflows are superposed along the line of sight at their launching point because of the beam smearing effect, hence, the cancellation between positive and 
negative velocities leads to net zero velocity, while the broad velocity distribution results in high velocity dispersion \citep[e.g.,][]{Durre2019,Shin2019b}. 
Therefore, our scenario of a 2nd AGN can naturally explain the gas kinematics at X1. 
Second, the morphological symmetry (i.e., the elongated tail-like structures) between R1 
and B1 in the velocity map suggests that they are physically related, sharing a same origin (i.e., the 2nd AGN). 
Third, we detect the radial velocity pattern of acceleration followed by deceleration for both of the gas blobs R1 and B1, which are centered 
at the potential launching position X1 (see Figure 3). This increasing and decreasing velocity pattern is a signature of gas outflows detected in a number of nearby AGNs
\citep{Crenshaw2000a, Crenshaw2000b, Muller-Sanchez2011,Durre2019}. While the radial velocity pattern, as well as the relatively 
high gas velocity  in NGC 1068 were already noticed in previous studies \citep{Crenshaw2000a,Muller-Sanchez2011}, 
gas outflows are assumed to be launched from the nucleus. 
Fourth, we find a biconical shape centered at X1 in the \OIII/\Hb\ ratio map, which is presumably due to an ionizing source at X1 (i.e., the 2nd AGN). Note that the approaching gas blob B1 was considered to be driven by the 1st AGN at the nucleus in the previous studies \citep[e.g.,][]{Cecil2002,Das2006, Gerssen2006}. 
However, the gas blob B1 is more extended than R1, covering the two slit positions shown in Figure~3, and it is likely that B1 is the combination of the approaching gas blobs, which are launched form the nucleus and the position X1, respectively (see the schematic view of two pairs of biconical outflows in Figure~6). 

\begin{figure*}{}
\centering
\includegraphics[width =  0.9\textwidth]{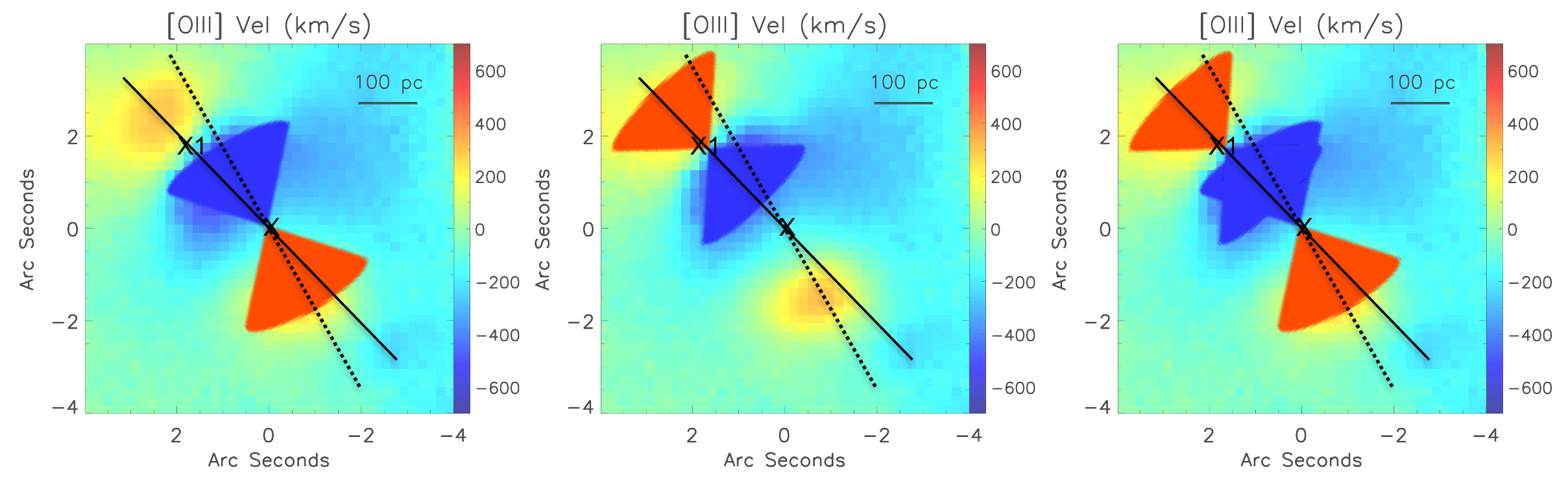}
\caption{
Schematic view of gas outflows in NGC 1068 overlaid on the velocity map of \OIII. Gas outflows launched from the nucleus, X (left, PA= 30$\arcdeg$), 
X1 (middle, PA= 44$\arcdeg$), and both of them (X and X1, right) are presented. Solid and dotted lines are the same as in Figure~3.\\
\label{fig:allspec1}}
\end{figure*} 

We investigate whether the kinematics of gas blobs R1 and B1 can be explained by a rotating disk, which is centered at X1. 
As shown in the lower panel of Figure~3, the peak velocity of B1 and R1 reaches $\sim$500 \kms\ at the projected 
distance of only $\sim$50 pc from X1. To produce a rotational velocity of 500 \kms\ at 50 pc, the required 
enclosed mass needs to be a minimum of $10^{9.5} \rm \ M_{\odot}$ when the disk is assumed to be edge on.
For the case of NGC 1068, however, the enclosed mass within the central 50 pc is only $10^{7.2} \rm \ M_{\odot}$, as estimated based on the stellar rotational velocity (i.e., $\sim$30 \kms\ at 50 pc from the nucleus) and the inclination 
(i.e., 40$\arcdeg$, \citealt{Bland-Hawthorn1997}). Because the expected mass of the gas disk at X1 is a 
factor of $\sim$200 higher than that based on stellar kinematics, a rotating gas disk is not likely responsible for the kinematics of gas blobs B1 and R1. Previous studies of nearby AGNs, including NGC 1068, pointed out the same conclusion that such high 
velocity (> 300\kms) of ionized gas is due to gas outflows, not a galactic disk rotation \citep[e.g.,][]{Muller-Sanchez2011,Durre2019}. 

Alternatively, supernova (SN) may be the origin of R1 as there was a report of
high-velocity molecular gas outflows ($V\sim$500\kms) driven by a SN in NGC 6240 \citep{Treister2020}.
However, SN is not likely responsible for R1, because of the following two reasons.
First, as shown in Figure~4, the major ionizing source for R1 is AGN \citep[see also ][]{DAgostino2019,Mingozzi2019}. 
While \cite{DAgostino2019} found significant shock contribution in the central region ($\sim$0.5 kpc$\times$1 kpc), 
it was interpreted to be the gas outflows launched from the 1st AGN of NGC 1068, but not due to other mechanisms (i.e., SN).
Second, an X-ray study by \cite{Ogle2003} showed that a region around X1 (i.e., 3\arcsec\ NE of the nucleus) 
is mainly photoionized rather than collisionally ionized, implying that AGN (not SN) is the major ionizing source of the region.
In fact, our BPT analysis clearly showed that the photoionization of R1 is dominated by AGN (see Figures 4 and 5).
Note that we cannot rule out the SN origin of R1 if the SN-driven photoionization is relatively weak and undetectable 
as the BPT map requires. In this case,  it is not clear how the SN provides a strong impact on the gas kinematics.

{ One may expect a light excess around X1 in the flux map, if there is indeed a mass-accreting black hole.
While we find no strong evidence, we detect an excess of \OIII\ flux around X1 in the flux map of ionized gas (see Figure~3). 
In contrast, there is no signature of dynamical disturbance in the stellar flux and kinematical maps, which may suggest 
that the galaxy-galaxy mass ratio is very large and the contribution of the merged component is negligible and not detected in the flux-weighted stellar absorption lines.}
The possibility of minor mergers in NGC 1068 has been discussed based on the detections of 
(1) ultradiffuse objects around NGC 1068 \citep{Tanaka2017}, (2) an off-centered circumnuclear disk 
\citep[CND,][]{Garcia-Burillo2014}, and (3) the non-Kerplerian motion of molecular gas in the molecular 
torus \citep[e.g.,][]{Imanishi2018}. We note that \cite{Garcia-Burillo2014} also discussed the presence of 
another AGN in the SW of the nucleus ($\Delta$RA=  --1.50\arcsec, $\Delta$Dec=  --1.00\arcsec)
to explain the off-centered CND even though it is yet to be observationally confirmed.

Recently, \cite{Wang2020} claimed a close binary supermassive black hole (SMBH) with a separation of $\sim$0.1 pc,
in order to account for a counter-rotating disc at the nucleus of NGC 1068. If the sub-pc scale binary as well as our proposed 2nd AGN are present, it means that NGC 1068 contains three SMBHs. While the observational evidence for triple SMBHs are scarce, 
a few triple SMBHs have been recently detected (e.g., NGC 6240, \citealt{Kollatschny2020}; and SDSS J0849+1114, \citealt{Liu2019,Pfeifle2019}). 
According to the cosmological simulation by \citet{Kelley2017}, triple SMBHs are expected to be up to 30\%\ among binary SMBHs,
suggesting that the presence of a triple SMBH in NGC 1068 is feasible although the confirmation is beyond the scope of this work.\\

\subsection{Counterpart in multiwavelength data}\label{Discussion}

To investigate whether the 2nd AGN scenario is consistent with other observational signatures, 
we use X-ray and radio data and search for evidence of multiple cores at the central region. 
Starting with X-ray data, we check the Chandra high-resolution camera (HRC) data of NGC 1068 (ObsID: 12705; PI: Fabbiano), which were previously 
presented by \cite{Wang2012}. As shown in the top-right panel of Figure~1 of \cite{Wang2012}, the Chandra 
image with the PSF deconvolution resolves another putative X-ray point source at 3.6\arcsec\ NE of the nucleus 
($\Delta$RA=  +2.00\arcsec, $\Delta$Dec=  +2.75\arcsec. The detection of the point source is significant at the >7$\sigma$ level
when we take into account the Poisson noise of the adjacent pixels. 
The position of the X-ray point source is close to the center of the pair of gas blobs R1 and B1 (i.e., X1 position), 
albeit with an offset of $\sim$0.9\arcsec\ to the NE of X1, 
and it is likely to represent the 2nd AGN around X1. The spatial offset can be explained by
various effects (i.e., dust obscuration and ISM impact), which can cause the positional shift of the kinematic center of 
the gas outflows (i.e., X1) from the position of an X-ray point source, where an AGN is located \citep[see e.g.,][]{Durre2018}.
For example, a $\sim$0.7\arcsec\ separation was found between the kinematic 
center of gas outflows and the position of the central X-ray point source in NGC 5728 (\citealt{Durre2019}; 
see also \citealt{Shin2019b}). 

To investigate the possibility of the X-ray point-source as a potential AGN, we measure the X-ray luminosity in 2--10 keV
by converting the X-ray photon counts (133) at the pixel of the X-ray point source position, using the PSF 
deconvolved Chandra HRC image. For this practice, we calculate a conversion factor from
{\tt WEBPIMMS}\footnote{http://cxc.harvard.edu/toolkit/pimms.jsp} using the effective area curve of $Chandra$ 
Cycle 12, 
a power law index ($\Gamma$ =  1.8), and a Galactic column density 
($N_{H} =  2.99 \times 10^{20} \ \rm cm^{-2}$, \citealt{Murphy1996}). 
As a result, we obtain the 2-10 keV X-ray 
luminosity as $\sim8.7\pm0.8\times10^{38}$ erg $\rm s^{-1}$. Note that the uncertainty is from the Poisson statistics.
 The X-ray luminosity is not sufficiently high to confirm the presence of an additional AGN
because X-ray binaries can be the origin of the X-ray emission. Note that the Eddington luminosity of a 10 \msun\ stellar black hole is $\sim$10$^{39}$ \ergs, while the luminosity of high-mass X-ray binaries is typically less than 10$^{39}$ - 10$^{40}$ \ergs\ \citep[e.g.,][]{Grimm2003,Fabbiano2006, Mineo2012}.

We investigate the possibility that the luminosity of the X-ray point source is underestimated as the potential 2nd AGN is likely to be type 2 without presenting broad emission lines in the optical spectrum. In this case, we do not directly measure the intrinsic luminosity because of the obscuration by a dust torus. 
It has been known for the 1st AGN in NGC 1068 that the transmitted X-ray emission from the central source is completely obscured by a dust torus as the column density of this Compton-thick source is $N_{\rm H}$ $\sim10^{25}$cm$^{-2}$. Thus, it is difficult to measure the intrinsic X-ray luminosity as we only observe the X-ray emission reflected by the inner wall of the dust torus and ionized gas \citep[][but see also \citealt{Marinucci2016}]{Colbert2002,Ogle2003,Matt2004,Bauer2015}. To determine the intrinsic X-ray luminosity, for example, \citet{Colbert2002} used the \OIII\ luminosity as a proxy for the bolometric luminosity of the central AGN, estimating the intrinsic X-ray luminosity of $\sim$3.2 $\times10^{43}$ \ergs, which is a factor of 250 higher than the reflected X-ray luminosity. 
More reliably, \citet{Bauer2015} performed X-ray spectral analysis, reporting the intrinsic X-ray luminosity of the central AGN as 2.2$\times10^{43}$ erg $\rm s^{-1}$ 
(see \citealt{Marinucci2016}). 

In the case of the potential 2nd AGN, we cannot use the same correction factor obtained for the 1st AGN because the 2nd AGN is not likely a Compton-thick source. Instead, a proper spectral analysis is required to determine the intrinsic X-ray luminosity. Note that the AGN X-ray spectrum is complex as it is composed of various components, including soft X-ray access and a power-law component. Due to the lack of spectral information on the X-ray source around X1, we estimate the intrinsic X-ray luminosity from the observed X-ray luminosity by assuming a typical column density for type 2 AGNs and calculate the correction factor with {\tt WEBPIMMS\footnote{https://heasarc.gsfc.nasa.gov/cgi-bin/Tools/w3pimms/w3pimms.pl}}. We obtain a range of the obscuration correction factor from 1.06 and 7.14, depending on the column density N$_{\rm H}$= 10$^{22}$ to N$_{\rm H}$= 10$^{24}$. As an example, if we multiply the correction factor of 1.56, corresponding to N$_{\rm H}$= 10$^{23}$, to the observed X-ray luminosity of $\sim8.7\times10^{38}$ erg $\rm s^{-1}$, the intrinsic X-ray luminosity is $1.3 \times10^{39}$ \ergs. 
While we cannot measure the exact correction factor for the X-ray source due to the lack of X-ray spectral information, it is likely that the intrinsic X-ray luminosity is not sufficiently high to rule out a high-mass X-ray binary as an origin.

In addition, we analyze Advanced CCD Imaging Spectrometer (ACIS) data (ObsID= 370, 0.4 s frame, exposure time= 11.5 ksec), which were
presented by \cite{Young2001}, in order to check whether the X-ray spectrum of the X-ray point source shows  a
power-law continuum, as a power-law continuum is another observational signature of AGNs. Note that 
we do not investigate deeper data (ObsID= 344, 3.2 s frame, exposure time= 47.4 ksec), which were heavily affected by pile-up 
at around X1 (and also the nucleus) because of the high count rate \citep[see ][]{Young2001}.
We extract the ACIS spectrum from the position of the X-ray point source as detected in the Chandra HRC image with a radius of 0.5\arcsec. Then, 
we fit the spectrum with a various combinations of 1) a single-temperature plasma model (MEKAL), 
2) bremsstrahlung, 3) power law, and 4) a thermal plasma component (APEC). All of the models include
the absorption by the Galactic column density ($N\rm_{H}(Gal)$ =  2.99 $\times\ 10^{20} \rm \ cm^{2}$, \citealt{Murphy1996}) 
as similarly done by \cite{Young2001}. However, we find that the inclusion of a power law does not improve 
the fitting results compared to those without a power law. 

To further investigate the origin of the X-ray point source, we measure the \OIII/soft X-ray ratio, which is a tracer of ionization states \citep{Bianchi2006}. By investigating the ratio for several X-ray knots in the central region of NGC 1068,
\cite{Wang2012} reported that a few of them were originated from shocks. Using the same data set and method of \cite{Wang2012}, we find the ratio of $\sim8.6\pm0.86$ at the pixel of the X-ray point-source position (i.e., within the extraction area of $0.13\arcsec \times 0.13\arcsec$). This result indicates photoionization rather than shock. 
{ Note that the flux ratios, \OIII/\Hb\ and \NII/\Ha, \SII/\Ha, or \OI/\Ha\ obtained around the X1 position are also 
consistent with AGN photoionization in the BPT diagrams. Thus, we conclude that X-ray emission is unlikely to originate 
from shocks unless \OIII\ flux is strongly dominated by the photoionization by the 1st AGN. }

Secondly, we check the ALMA data of CO (3-2) \citep{Garcia-Burillo2014}, and dense molecular gas 
(i.e., HCN (3-2) and HCO$^{+}$ (3-2)), presented by \cite{Imanishi2018}, as well as the VLA radio data by used by 
\citet{Wilson1983,Gallimore1996a}, while we do not find any strong
signature (i.e., radio core) of the 2nd AGN around X1. 
This can be explained with the low luminosity and/or the low mass of the host galaxy of the 2nd AGN, as discussed in the previous section.

Overall, we find no clear evidence of a second AGN based on the multiwavelength data because
the X-ray luminosity is relatively low and comparable to high-mass X-ray binaries. Nevertheless, 
we find no evidence to rule out the possibility of the 2nd AGN.\\

\subsection{Hidden broad emission line}\label{Discussion}
One of the characteristics of an AGN is the presence of broad emission lines (i.e., FWHM > 2000 \kms), which are emitted
from the sub-pc scale broad line region. For Seyfert 2 galaxies, the broad line region is obscured by a dust 
torus, but it can be detected through spectropolarimetric observations. As described in Section 1, NGC 1068 is the first 
type 2 Seyfert galaxy whose hidden broad Balmer lines were detected \cite[e.g.,][]{Antonucci1985}.
Since then, a number of spectropolarimetric results have been presented for NGC 1068. Among them, spatially resolved
spectropolarimetric studies (\citealt{Inglis1995}, see also \citealt{Miller1991}) found the interesting result that the line-width 
of the hidden broad Balmer lines detected in the NE (e.g., $\sim3250\pm400$ \kms\ at 5\arcsec, or 360 pc NE) of the 
nucleus is significantly smaller than that detected at the nucleus ($\sim4400\pm300$ \kms). \citet{Inglis1995} presented 
the line width measurements of the hidden broad \Ha\ line, which were detected at multiple points from the nucleus to the NE direction 
(i.e., 2.5, 5, and 7.5\arcsec\ offset, corresponding to 180, 360 and 540 pc distance from the nucleus), 
show a consistency ($\sim$3000 \kms) within 1-$\sigma$ (see the upper panel of Figure~7).
This result may suggest that the hidden broad \Ha\ detected in the NE has a different origin from
that at the nucleus.
One possible explanation is that there is an additional AGN in between 2.5--7.5$\arcsec$ (or 180--540 pc) NE of the nucleus, 
whose hidden broad lines are intrinsically narrower than those of the 1st AGN at the center because of the smaller black hole mass of the 2nd AGN.

\begin{figure}{}
\centering
\includegraphics[width =  0.43\textwidth]{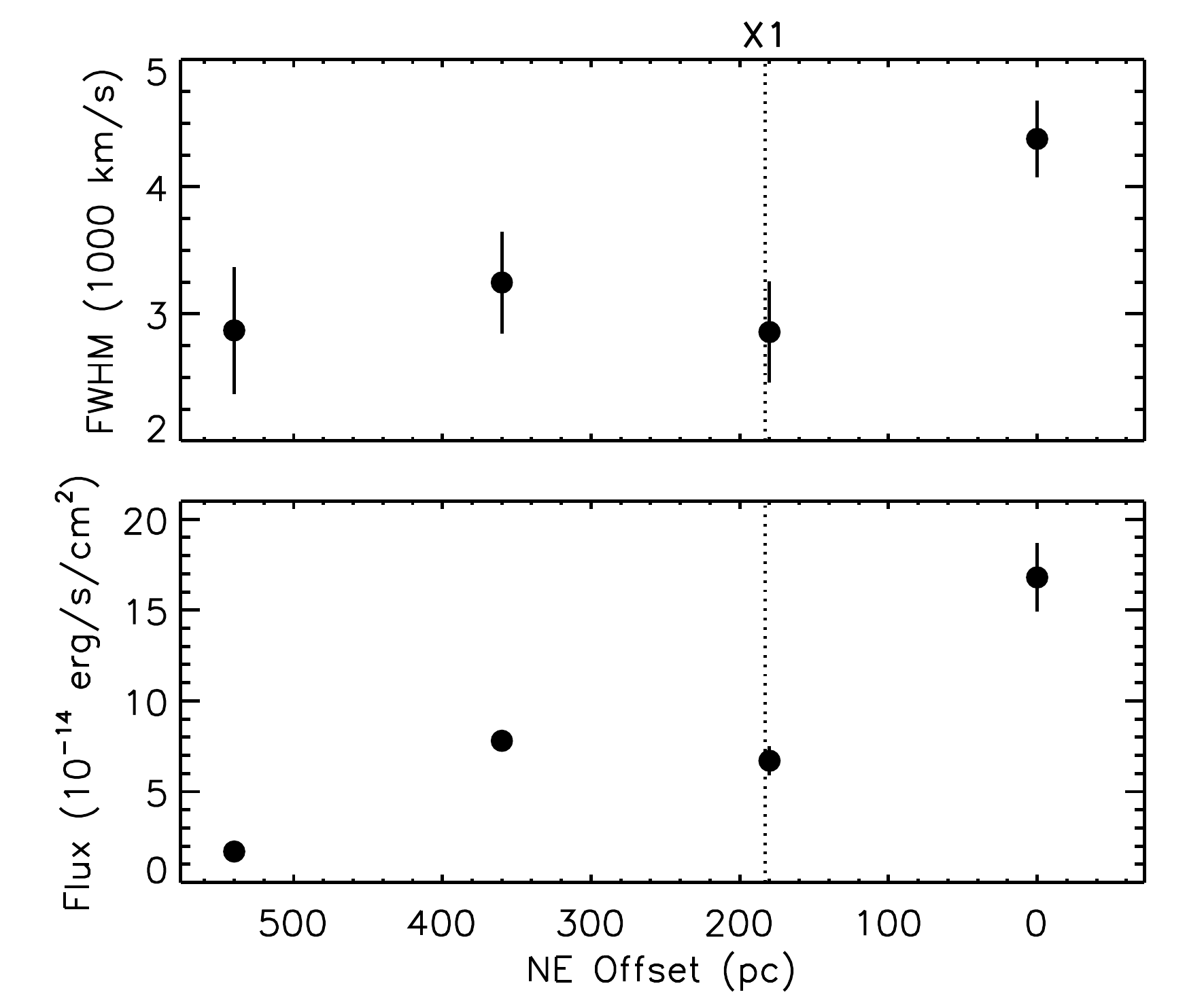}
\caption{
One-dimensional distribution of the FWHM (upper) and line flux (lower) of the hidden broad component of \Ha. 
A dotted line denotes the position of X1. All measurements are adopted from \cite{Inglis1995}. \\
\label{fig:allspec1}}
\end{figure} 

As shown in the lower panel of Figure~7, the line flux of the hidden broad \Ha\ line rather increases at 
360 pc NE of the nucleus, which is in disagreement with the trend of radially decreasing flux if it originated from the 
nucleus. This trend is consistent with the scenario of a 2nd AGN
at a distance of 360 $\pm$ 90 pc NE from the nucleus. Note that the kinematic center of R1 and B1 is X1, 
which is positioned at 180 pc from the nucleus; however, the true location of an AGN can be somewhat offset from 
the kinematical center as discussed in Section 4.2.
It is worth noting that imaging polarimetric observations have detected a highly polarized knot at 4.7\arcsec\ 
NE of the nucleus \citep{Miller1991,Scarrott1991,Simpson2002}. It could be also associated with the second 
AGN while it is 2\arcsec\ offset from X1.

In previous works, the origin of the broad Balmer emissions and high polarized intensity in the NE region was 
interpreted as the scattered light from the nucleus \citep[e.g.,][]{Miller1991,Inglis1995} and molecular gas, which 
prevents the expansion of the radio jets \citep{Simpson2002}, respectively. However, these scenarios cannot clearly 
explain the difference of the line widths in the NE region as well as the bump in
the flux distribution at 360 pc. In contrast, the scenario with the 2nd AGN positioned at around X1 may provide a better explanation. \\

\subsection{Mass outflow rate and Energetics}\label{Discussion}

We discuss whether the energetics of the potential 2nd AGN is eligible to drive gas outflows manifested by gas blobs B1 and R1,
by investigating the mass outflow rate and kinetic energy output. We use two different methods, namely, geometric and 
luminosity approaches \citep[see e.g.,][]{Revalski2018}.

\subsubsection{Geometric approach}
For the geometric approach, 
the mass outflow rate and energy budget can be calculated as follows \citep[see, e.g., the equation 1 of][]{Muller-Sanchez2011},
\begin{equation}
\dot{M}_{\mathrm{out}} = 2\, m_\mathrm{p}\, N_e\, v_{\mathrm{max}}\, A\, f
\end{equation} 
\begin{equation}
\dot{E}_{\mathrm{out}}=1/2\,\dot{M}_{\mathrm{out}}\,(v_{\mathrm{max}}^2+\sigma^2)
\end{equation} 
This approach is based on a geometric assumption by  
determining five parameters: (1) the electron density, $N_{e}$, (2) the deprojected maximum outflow velocity, $v_{max}$, (3) the lateral surface area of outflows at the maximum velocity position, $A$, (4) the averaged velocity dispersion in gas outflows, $\sigma$, and (5) the filling factor, $f$.  
Because we interpret B1 as a superposed outflows, respectively launched from X and X1, we only use R1 to calculate outflow parameters. 
First, we estimate the electron density based on the \SII$\lambda$6717/\SII$\lambda$6731 ratio, which is
measured from a flux-weighted spectrum extracted within a 1\arcsec\ radius from the maximum velocity position in R1. 
By assuming the electron temperature to be 10000 K, we obtained an electron density of 907 $\rm cm^{-3}$, which is consistent within 
$\sim$10\%\ with the previously reported value \citet[][i.e., $\sim$800 $\rm cm^{-3}$]{Mingozzi2019}. 
Second, we adopt the (projected) maximum velocity of \OIII\ in R1 (i.e., 291 \kms) as the lower limit from the velocity map
(see Figures 2 and 3) because the inclination of the outflows is unknown.  
Third, we estimate the lateral surface by determining the size of the major axis of R1. 
If outflows have a biconical shape, the lateral surface would be round, which is shown as an ellipse
in the plane of the sky. Thus, the size of the major axis can be used to calculate the area of a circle. 
Based on this idea, we determine the size of the major axis to be 2\arcsec\ (i.e., 142 pc) by drawing a line through the maximum velocity position in R1, which is perpendicular to the gas outflow direction (i.e., Slit 1; see Figure~3).
To account for the hollow shape of outflows \citep[e.g.,][]{Muller-Sanchez2011,Bae2016,Shin2019b}, we adopt the hollow cone geometry of the 1st gas outflows, which were constrained as outer angle= 27$\arcdeg$ and inner angle= 14$\arcdeg$) based on biconical gas outflow modeling by \cite{Muller-Sanchez2011}.  
Using these estimates, we calculate the lateral surface area as 1.3 $\times 10^{4} \rm \ pc^{2}$. 
Fourth, the average velocity dispersion in R1 is measured as 485 \kms using the same spectrum used for electron density estimation. 
Finally, we adopt the filling factor of 0.11 from \cite{Storchi-Bergmann2010}, who calculated the filling factor for gas outflows 
in NGC 4151. 
Because we adopt the projected maximum velocity, these parameters provide the lower limit of the mass outflow rate and energy outflow rate as $\dot{M}_{\rm out}$ = 18.1 M$_{\odot}$ yr$^{-1}$ and $\dot{E}_{\rm out}$ = 1.9$\times$10$^{42} \rm \ erg  \ s^{-1}$, respectively. 

For a consistency check, we calculate the energetics of the gas outflows (R2) driven by the 1st AGN. 
Applying the same method to R2, we measure the averaged electron density in R2 (479 $\rm cm^{-3}$), 
the maximum velocity (321 km s$^{-1}$), the lateral surface area (0.8 $\times 10^{4} \rm \ pc^{2}$), 
and the averaged velocity dispersion in R2 (544 \kms). By adopting the same filling factor of 0.11, we obtain the lower limit of the 
mass outflow rate as 6.5 M$_{\odot}$ yr$^{-1}$ and the energy outflow rate as 8.2$\times$10$^{41}$ erg $\rm s^{-1}$. If we adopt the inclination of the 
1st gas outflows (i.e., 9$\arcdeg$) presented by \cite{Muller-Sanchez2011}, the deprojected (intrinsic) maximum velocity 
becomes 2057 km s$^{-1}$, which is consistent within 10\%\ with that constrained by \cite{Muller-Sanchez2011}. 
We find that the mass outflow rates and kinetic energy are $\dot{M}_{\rm out}$ = 38.8 M$_{\odot}$ yr$^{-1}$ 
and $\dot{E}_{\rm out}$ = 4.8$\times$10$^{43}$ erg $\rm s^{-1}$, respectively, which are consistent within 
an order of magnitude with those reported by \cite{Muller-Sanchez2011}.\\ 

\subsubsection{Luminosity approach}
The luminosity approach calculates mass outflow rate and energy budget as follows,
\begin{equation}
\dot{M}_{\mathrm{out}}  =  3 M_{\mathrm{gas}} \frac{v_{\mathrm{out}}}{R_{\mathrm{out}}}
\end{equation}
\begin{equation}
\dot{E}_{\mathrm{out}} =  \frac{1}{2} \dot{M}_{\mathrm{out}} v^2_{\mathrm{out}},
\end{equation} 
where $M_{\mathrm{gas}}$ is the ionized gas mass, $v_{\rm out}$ is the averaged outflow velocity 
(i.e., $v_{\rm out}$=$\sqrt{v_{\rm gas}^{2} + \sigma_{\rm gas}^{2}}$) within the outflows, and $R_{\mathrm{out}}$ is the outflow size \citep[e.g.,][]{Karouzos2016b,Bae2017,Kang2018}.
First, the ionized gas mass is converted from the \OIII\ luminosity with an estimated electron density 
using the equation of $M_\mathrm{gas} = 0.4 \times 10^8 M_{\odot} \times (L_{\mathrm{[O\, III],43}}) (100 \, \mathrm{cm^{-3}}/n_e)$
\citep[see, e.g.,][]{Carniani2015,Karouzos2016b}.
In the previous works by our group \citep[e.g.,][]{Karouzos2016b,Bae2017},
the ionized gas mass was estimated using the integrated \OIII\ luminosity within the outflow size. 
{ If NGC 1068 hosts an additional AGN along with its corresponding gas outflows, 
it is not trivial to measure the \OIII\ emission solely due to the photoionization by the 2nd AGN.
For simplicity, we constrain the upper limit of the ionized gas mass by combining \OIII\ flux around X1. We calculate twice the integrated \OIII\ luminosity in R1 (4.6$\times10^{40} \ \rm erg \ s^{-1}$) within a 1\arcsec\ radius as the \OIII\ luminosity
and calculate the ionized gas mass to be M$_{\rm gas} = 2\times10^{4} \rm \ M_{\odot}$ using the average electron density in this region (907 $\rm cm^{-3}$). 
Note that assuming a bicone, we multiply a factor of 2 to the \OIII\ luminosity of R1. Because a large fraction of the warm gas is photoionized by the 1st AGN, the derived gas mass should be considered as an upper limit. While the \OIII\ map in Figure 2 shows a dominance of photoionization by the 1st AGN even at the R1 area, we detect additional flux at X1 and R1 areas as shown in Figure 3, which is presumably due to the photoionization by the 2nd AGN. While it is difficult to estimate the fraction of the ionized gas flux due to the 2nd AGN with respect to the total \OIII\ flux in R1 area, the fraction should not be negligible because we clearly detect distinct kinematical signatures in the velocity and velocity dispersion maps based on the flux-weight \OIII\ line profile. If we assume the \OIII\ flux due to the 2nd AGN is $\sim$1-10\%, the total gas mass should be corrected accordingly. 
Second, the average outflow velocity is calculated using the average velocity (232 \kms) and velocity dispersion (485 \kms) in R1. 
Third, we determine the outflow size as 140 pc (or $\sim$2\arcsec), which is estimated at the position 
where the \OIII\ velocity approaches 100 \kms (see $\sim$--330 pc in the middle panel of Figure~3). Note that
the size of outflows is uncertain, and we simply use this value as our best approximation.
As a result, we obtain the upper limit of the gas mass and energy outflow rate as $\dot{M}_{\rm out}$ = 0.24 M$_{\odot}$ yr$^{-1}$. 
and $\dot{E}_{\rm out}$ = 2.2$\times$10$^{40}$ erg $\rm s^{-1}$. The actual gas mass and energy outflow rate should be corrected by
the fraction of the \OIII\ flux due to the 2nd AGN over the total \OIII\ flux. 
}

\subsubsection{Comparison of outflow energetics with bolometric luminosity}
Based on the two aforementioned methods, we estimate the mass outflow rate and energy outflow rate. However, there is a large discrepancy between them
by $\sim$2 orders of magnitude, indicating the systematic uncertainties of the outflow energetics. \citet{Harrison2018} pointed out that there are large differences in the
mass outflow rate and energy outflow rate among various studies due to the different analysis methods and inhomogeneous definitions of the outflow size and velocity. 
\cite{Revalski2018} also discussed the huge difference between the outflow rates measured from the two methods and
recommended using the luminosity approach because it better traces the physical parameters 
of gas outflows (i.e., ionized gas mass).
While the derived mass outflow rate and energy outflow rate are within the range of the reported values in the literature  \citep[e.g.,][]{Greene2011,Harrison2014,Karouzos2016b,Bae2017, Rakshit2018}, we emphasize that there are large systematic uncertainties in determining outflow energetics.

Nevertheless, we compare the derived outflow energetics with AGN bolometric luminosity to investigate the kinetic coupling efficiency (i.e.,  $\dot{\rm E}_{\rm out}$/L$_{\rm bol}$). To estimate the bolometric luminosity of the 2nd AGN, we utilize two different indicators, X-ray luminosity and \OIII\ luminosity. If we use the luminosity of the X-ray source around X1
with an obscuration correction factor of 1.56 for $N_{\rm H}$= 10$^{23} \ \rm cm^{-2}$ and a bolometric correction factor of 10.85, we obtain the bolometric luminosity of the 2nd AGN as L$_{\rm bol}$= 1.5$\times$10$^{40}$ \ergs. Here the bolometric correction factor is calculated based on the Equation 2 of \cite{Duras2020} with the coefficients for type 2 AGNs. 
Note that L$_{\rm bol}$ could be lower by a factor of $\sim$1.5 or higher by a factor of $\sim$7, if $N_{\rm H}$ is 10$^{22} \ \rm cm^{-2}$ or
10$^{24} \ \rm cm^{-2}$, respectively. 
Also, note that the bolometric correction for the hard X-ray is relatively uncertain. On the other hand, if we use the upper limit of \OIII\ luminosity around R1,
along with a bolometric correction of 3500 \citep{Heckman2004}, we derive the upper limit of L$_{\rm bol}$= 1.6$\times$10$^{44}$ \ergs.
If we assume 1\% of the total \OIII\ in R1 is photoionized by the 2nd AGN, we obtain L$_{\rm bol}$ = 1.6$\times$10$^{42}$ \ergs.
Note that the large difference between X-ray based and \OIII\ based bolometric luminosities is mainly due to various uncertain factors, including the obscuration correction factor and bolometric correction of hard X-ray luminosity as well as the uncertain fraction of the \OIII\ flux due to the 2nd AGN around R1 area. 

Comparing with these estimates of bolometric luminosity, the energy outflow rate determined by the geometric approach is challenging because
the kinetic energy carried by the outflow is much larger than the X-ray based bolometric luminosity, with the kinetic coupling efficiency of $\sim$130.
In contrast, if we adopt the OIII-based bolometric luminosity, the efficiency is $\sim$1.2.
On the other hand, the energy outflow rate based on the luminosity approach is at least two orders of magnitude lower than that of the geometric approach, and the kinetic coupling efficiency becomes substantially smaller. For example, if we assume a 1\%\ of the total \OIII\ flux in the R1 area is due to the 2nd AGN, we obtain $\dot{E}_{\rm out}$ = 2.2$\times10^{38}$, which is $\sim$0.015 and 
10$^{-4}$ of the X-ray based and \OIII-based bolometric luminosities, respectively. 
These results are within the range of the kinetic coupling efficiencies of warm ionized gas reported in the literature \citep{Harrison2018}.
Albeit with large systematic uncertainties of outflow energetics and AGN bolometric luminosity due to a number of assumptions,
we find no strong evidence to rule out the scenario of outflows driven by a 2nd AGN.

\section{Conclusion}

In this work, we investigate the complex kinematics of ionized gas at the central region of NGC 1068.
The main results are summarized below. \\
 
$\bullet$ We detect a pair of blueshifted and redshifted gas blobs, which are located in the NE region at 
$\sim$110 and $\sim$250 pc distance, respectively, from the nucleus. The spatially-resolved kinematics of the 
two gas blobs are similar but inconsistent with a rotation or simple biconical outflows launched from the central AGN.  \\

$\bullet$ The center of the two gas blobs at 180 pc distance from the nucleus is characterized by zero velocity and 
high velocity dispersion to the line-of-sight, which are typical kinematical features of a launching point of AGN-driven gas outflows. 
Also, the pair of the gas blobs shows kinematically similar morphology in the velocity
map and shares a consistent radial velocity structure with an increasing and decreasing trend, 
implying that these outflows are launched by an additional (2nd) AGN, which is located at the midpoint of the pair.\\

$\bullet$ High resolution X-ray data shows a putative X-ray point source near the expected position of the 2nd AGN, while its low 
luminosity is not sufficient to confirm the 2nd AGN { or rule out the X-ray binary origin}.\\

$\bullet$ Based on the spatially resolved spectropolarimetric analysis, we find in the NE region a bump in the flux distribution and a significantly 
smaller velocity dispersion of the polarized broad Balmer line compared to that of the nucleus. 
These results  are consistent with the scenario that there is an additional AGN at the midpoint of the two gas blobs. \\

Based on multiwavelength analysis (i.e., optical and X-ray), we provide circumstantial evidence for the presence of an additional AGN.
However, the analysis results are insufficient to confirm the case and there are other possibilities, i.e., X-ray binary and SN shock, responsible for the complex kinematics of ionized gas.
One way to confirm the 2nd AGN scenario is to use spectropolarimetric observations with a high spatial resolution to separate the central and additional AGNs. 
In turn, if the 2nd AGN is confirmed, the kinematical characteristics such as zero velocity and high velocity 
dispersion of ionized gas may be used as an alternative way to search for multiple AGNs. 
\\

\acknowledgements
We thank the anonymous referee for comments that have improved the clarity of this paper.
We thank Bernd Husemann and Masatoshi Imanishi for useful discussions.
This research was supported by Basic Science Research Program through the National 
Research Foundation of Korea (NRF) funded by the 
Ministry of Education (2016R1A2B3011457 and 2019R1A6A3A01093189). 
JS and MK were supported by the National Research
Foundation of Korea (NRF) grant funded by the Korea government (MSIT)
(No. 2020R1A2C4001753). 
JW acknowledges support by the National Key R\&D Program of 
China (2016YFA0400702) and NSFC grant U1831205.
Based on observations collected at the European 
Southern Observatory under ESO programme 094.B-0321(A).

\end{document}